# 七论以用户为中心的设计：从自动化飞机驾驶舱到智能化飞机驾驶舱

## 许 为 *

（浙江大学 心理科学研究中心，杭州 310058）

**研究要点：**
1. 概述自动化飞机驾驶舱的人因问题和基于"以人为中心自动化"理念的人因工程研究
2. 针对智能化飞机驾驶舱，综述基于"以人为中心 AI"理念的人因工程初步研究，并展望未来研究
3. 根据智能人机交互的人因工程模型和协同认知生态系统框架，提出针对大型商用飞机单人飞行操作的人因工程初步解决方案

**摘要** 本文从"以用户为中心的设计"理念出发，在概述大型商用飞机自动化驾驶舱的人因问题和基于"以人为中心自动化"理念的人因工程研究基础上，综述基于"以人为中心人工智能"理念的智能化飞机驾驶舱人因工程初步研究，并且展望未来的人因工程研究。根据作者提出的智能人机交互人因工程模型以及协同认知生态系统框架，提出针对大型商用飞机单人飞行操作的人因工程初步解决方案，并且展望今后的人因工程研究。

**关键词** 以用户为中心的设计　自动化飞机驾驶舱　智能化飞机驾驶舱　协同认知系统　人因工程

**中图分类号**：B849　**文献标识码**：A　**文章编号**：

## User Centered Design（VII）:
## From Automated Flight Deck to Intelligent Flight Deck

### XU Wei

(Center for Psychological Sciences, Zhejiang University, Hangzhou 310058, China)

### Abstract

Driven by the "user-centered design" philosophy, this paper first outlines the human factors problems of the flight deck automation for large civil aircraft and the human factors research carried out based on the "human-centered automation" approach. This paper then reviews the previous initial human factors research on intelligent civil flight deck based on the "human-centered AI" approach and discusses the prospects for future human factors research. Based on our proposed human factors engineering model for intelligent human-computer interaction and the framework of joint cognitive eco-systems, this paper proposes an initial human factors solution for the single-pilot operations of large commercial aircraft and presents preliminary suggestions for future human factors research.

---

*作者：许为，博士，研究员；e-mail: xuwei11@zju.edu.cn。

1. 引言

大型商用飞机驾驶舱是一个典型的复杂人机系统，其机载人机交互技术经历了机械化、自动化的发展过程，目前正在逐步走向智能化。该复杂人机系统包括了人因工程（human factors engineering）研究和应用的几乎所有领域的内容（许为，陈勇，2012；许为，葛列众，2018）。

从发展角度看，机载人机交互技术最初的开发设计并非是遵循"以人为中心"的人因工程设计理念（冯志祥，白昀，2021；许为，2005）。早期的机械化飞机驾驶舱基本上遵循"人适应于机器"的设计理念。计算机技术带来了自动化飞机驾驶舱，但是最初的自动化驾驶舱设计基本上遵循"以技术为中心的自动化"理念(Billings, 1996)。自动化技术一方面带来了许多好处，同时也带来了一些人因问题，甚至间接或直接地造成了多起飞机解体的重大事故 (Endsley & Kiris, 1995; Mumaw et al., 2000)。进入智能时代，基于人工智能（AI）技术的智能系统逐步进入人们日常的工作和生活，大型商用飞机驾驶舱也不例外 (Bailey et al., 2017)。 如何基于"以人为中心"的设计理念，吸取自动化驾驶舱设计中的一些历史教训，利用智能新技术进一步提升商用飞机和乘客的安全，这是摆在人因工程面前的一个重要课题 (Billings, 1996; 许为，2004)。

本文首先概述针对大型商用飞机自动化驾驶舱的人因工程研究，然后讨论和分析智能化飞机驾驶舱的概念、人因工程研究的初步进展以及今后的发展。最后，应用作者所提出的智能人机交互的人因工程模型和协同认知生态系统框架（许为，2022），讨论和分析大型商用飞机驾驶舱单人飞行操纵模式中的人因问题，为今后这方面的工作提出人因工程的初步解决方案。

2. 自动化飞机驾驶舱

2.1 基本概念

自动化系统通常依赖于固定的逻辑规则和算法来执行定义好的任务，并产生确定的操作结果。自动化操作需要人类操作员启动、设置控制模式以及编制任务计划等，在一些特殊操作环境中（如设计无法预测的非正常操作场景或应急状态），需要人工干预来控制系统的运行。

基于自动化技术的现代大型商用飞机驾驶舱中的自动化飞行系统主要由飞行管理计算机系统和控制显示装置、自动驾驶仪和自动驾驶模式控制板、自动油门、飞行指引仪、自动驾驶模式指示仪、水平和垂直导航状态显示器等组成(见图1、2)。根据操作需求，飞行员不仅可以选择不同的自动化水平（手动、半自动、全自动），还可以对由自动油门、俯仰和横滚三个飞行维度组合而成的数十种自动化飞行控制方式作出选择。基于飞行员对飞行管理计算机系统所输入的指令和飞行规划，在正常飞行场景下，自动化飞行系统可以自动操控所有的飞行任务（除了起飞），当出现设计无法预料的非正常飞行场景时，需要飞行员通过手动飞行操纵来执行人工干预。

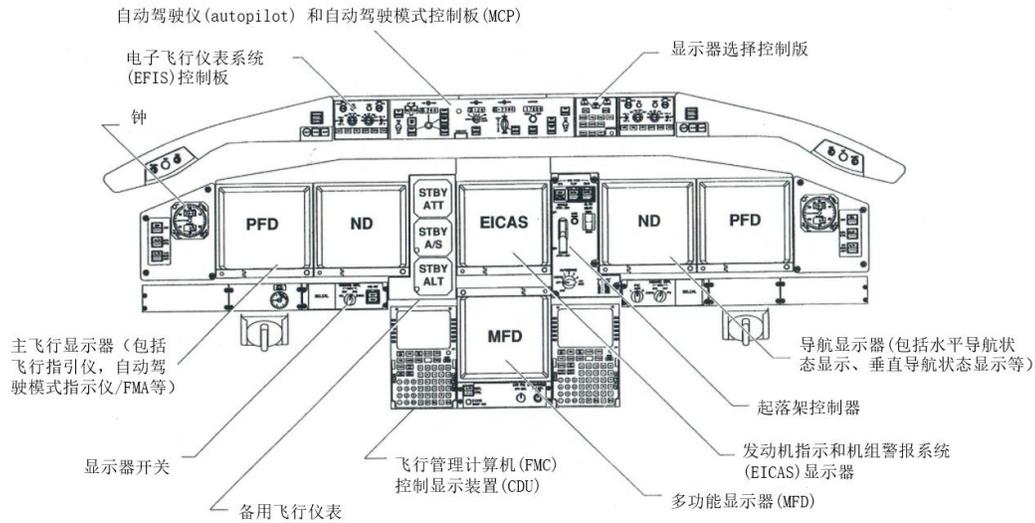

图 1   波音 777 飞机驾驶舱人机界面示意图（Boeing，2002）

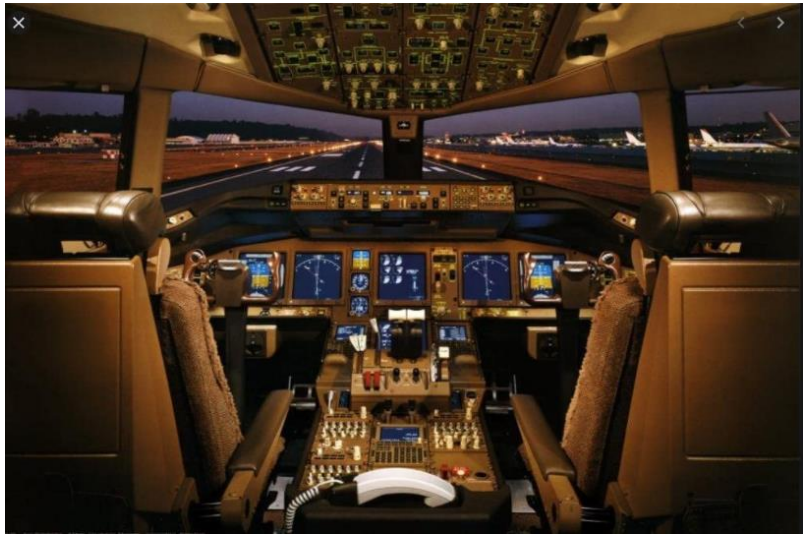

图 2   波音 777 飞机驾驶舱布局

**2.2 主要人因问题**

　　自动化驾驶舱提高了飞行操作的准确性及可靠性、经济性、乘客舒适性以及飞行安全，同时也改变了驾驶舱人机界面的设计、飞行操作方式以及操作程序，并且对飞行员能力、知识、操作、培训等方面提出了新要求(许卫，2004)。

　　在传统的非自动化飞机驾驶舱中，通过对各种仪表数据的判读和理解，飞行员手动完成各种飞行操作任务。根据 Rasmussen (1983) 认知工程的行为层次模型（SRK），这是一种数据-驱动型作业，表现出以技能式(即手动操控)行为为主，并带有一些规则式行为（即根据操作场景来选择合适的操作程序）以

及少量的知识式行为（即在应急状态中，飞行员根据拥有的领域知识来采取有效的操作策略）。在自动化飞机驾驶舱操作中，飞行员的操作更多的是通过显示器、控制显示装置和自动驾驶模式控制板等机载设备来实施对自动化飞行的监控，飞行员以上3种行为之间的相对比例正好与传统驾驶舱中的比例相反，即飞行员的操控作业更多地归属于规则式和知识式行为，有效的自动化监控操作更多地取决于飞行员对领域知识（自动化系统基本原理和飞行操作程序等）的理解；当遇到非正常飞行操作情景时，飞行员需要进行一定的推理决策（Sarter, Wickens et al., 2003；许为, 2003）。这种规则和知识-驱动式监控操作对飞行员的能力提出了更高要求，因此自动化驾驶舱虽然减轻了飞行员的体力工作负荷，但是增加了他们的认知工作负荷（Grubb, et a., 1994）。

在过去的几十年中，人因工程界针对现代大型商用飞机驾驶舱中的人-自动化交互方面的问题开展了广泛深入的研究，评估自动化驾驶舱对操作员的情境意识、自动化模式识别、警戒水平、信任、工作负荷、工作绩效等方面的影响，并且已达到基本一致的共识（例如，Sarter & Woods, 1995；Endsley, 2017a；许为, 2003）。例如，自动化操作中的监控作业会导致飞行员操作员警戒水平降低 (Hancock, 2013)、对自动化的过度信任（自满）、过度依赖自动化（Parasuraman & Riley, 1997）等人因问题。

研究发现许多自动化系统存在脆弱性，它们在设计所规定的操作情况下系统运行良好，但是在遇到意外事件需要人工干预的应急状态下，自动化系统可能导致操作员的"自动化惊奇"（automation surprise)反应 (Sarter & Woods, 1995)：即操作员可能无法理解自动化系统正在做什么，为什么要这样做，为什么它会从一种自动化模式转换到另一种模式。"自动化惊奇"可能引起飞行员的模式混淆、自动化情景意识下降、"人在环外"效应、诊断问题准确性和效率降低等人因问题，导致飞行员员可能会作出错误的人工干预，从而带来产生人为差错的隐患（Endsley, 2015； Endsley & Kiris, 1995；Wickens & Kessel, 1979； Young, 1969）。FAA航空安全报告系统(ASRS)的数据库显示，在1988～1995年期间，大约有105份事故报告与自动化驾驶舱有关。这些问题直接或间接地导致了一些重大飞机解体事故的发生(Endsley & Kiris, 1995; Mumaw et al., 2000)。

### 2.3 人因问题产生的主要原因

美国联邦航空局（FAA）的人因工程研究（1996）确认了导致以上这些人因问题的一系列原因，其中包括飞行员-自动化交互、人机界面设计、飞行员培训等。近期，Read 等（2020）分析了美国国家运输安全局（NTSB）、加拿大运输安全委员会（TSB）和澳大利亚运输安全局（ATSB）的事故数据库中16份涉及自动化系统的事故调查报告（1997年1月至 2018年6月），结果表明多种相互作用的因素导致了与自动化相关的事故，其中最主要的因素包括飞行员情境意识、飞行操作程序、飞行员决策、自动化设计以及航空公司制定的自动化操作规定。导致自动化驾驶舱人因问题的原因往往不是单一的（Lee, 2018；Kaber, 2018）。例如，飞行员情景意识的下降往往是由多种因素造成的，其中包括人机界面设计、飞行员培训和飞行操作程序等。以下概述人因问题产生的一些主要原因。

### 2.3.1 人机界面

目前驾驶舱自动化人机界面是经过多次技术升级的产物。大型商用飞机驾驶舱自动化系统的开发始于上世纪70年代，它的初期研发遵循了"以技术为导向"的理念（Billings, 1996）。这种开发方式为飞行员提供了几十种自动化模式和控制方式，导致了过分复杂的人机界面（参见图1、2）。例如，针对自动化垂直导航飞行操作，系统提供了垂直速度、飞行高度改变、垂直导航航路、垂直导航速度以及飞行路径角等众多的控制方式，增加了飞行员的认知工作负荷和人为差错的发生概率（例如自动化方式选择错误）（FAA, 1996；Xu, 2007）。

新增的驾驶舱自动化系统功能与已有功能在人机界面的整合设计方面也存在不足。尽管航空界一直在努力优化自动化人机界面的设计，但是受设计兼容性、飞行员培训成本、适航认证风险等因素的影响，最初的"以技术为导向"设计给后续新机型人机界面设计的改进带来许多挑战，导致自动化人机界面并

没有得到根本性的改进。新增的机载设备一方面为飞行安全提供了进一步的保障，但是也有可能进一步增加了人机界面的复杂性（Dismukes et al., 2007）。例如，驾驶舱机载系统为飞行员提供了多通道告警信号（听、视、触觉等），然而逐步增加的新机载系统（例如地形回避和警告系统）的告警信号并没有与原有告警信号系统实现有效的整合设计，有时会给高负荷应急状态下的飞行员造成信息过载，从而带来产生人为差错的隐患(许为，陈勇，2014)。

### 2.3.2 自动化水平

系统的自动化水平会影响操作员的工作绩效、工作负荷以及情境意识（Kaber, 2018）。Bainbridge (1983) 在总结了以往研究的基础上提出了一个经典的现象：自动化的讽刺（ironies of automation）：即自动化程度越高，操作员的介入越少，操作员对系统的关注度就越低；在应急状态下，操作员就越不容易通过人工干预来安全有效地操控系统。Onnasch 等（2014）针对 18 项自动化研究的元分析发现，自动化水平的增加有利于操作员工作绩效的提升和工作负荷的下降，但是操作员的情境意识和手动技能也下降。研究表明采用适当的自动化有助于避免这些人因问题的产生。例如，Endsley & Kiris（1995）的研究证明全自动化所导致的情境意识损失可以通过使用中等程度的自动化设计来弥补。

### 2.3.3 飞行员培训

大多数的飞行员培训内容主要侧重于自动化系统的基本操作知识，缺乏对如何在各种飞行场景中（包括非正常场景）选择合适的自动化水平、方式等方面的任务导向型知识和自动化飞行领域知识等方面的培训(Mumaw, Boorman et al., 2000)。从人因工程角度看，这些知识对于飞行员正确判断自动化系统的工作状态、建立完整有效的心理模型、提升规则和知识-驱动式监控操作能力、提高解决问题和决策能力是非常重要的（Billman et al., 2020）。

### 2.3.4 文化因素

文化因素会影响人的自动化操作作业和对自动化的信任（Chien et al., 2018）。对自动化驾驶舱来说，国家民族文化和航空公司企业文化对飞行员的自动化飞行操作会产生影响。例如，美国得克萨斯大学对 12 个国家的 5879 名飞行员的问卷调查研究表明，对于驾驶舱自动化的态度和使用，不同国家文化间的差别明显大于同一国家内不同航空公司企业文化之间的差别。其中，大部分飞行员认为，在高负荷工作状态下，应避免对飞行管理计算机系统的再编程，但是这种选择的差异范围在不同国家之间达到 35%～64%；而对于所在航空公司是否要求飞行员使用自动化飞行的问题，不同国家间的选择范围则达到 32%～84%之多（许卫，2004）。

### 2.3.5 飞行操作程序

Mumaw, Boorman 等（2000)的研究表明，当飞行员应对非正常飞行状态、执行空中交通管制(空管)请求时，自动化飞行操作程序并不能有效地支持飞行员的任务。在某些情况下，没有现成的操作程序来帮助飞行员处理这些情况。当飞行员的自动化心理模型和人机界面设计存在缺陷时，缺乏有效的飞行操作程序会增加飞行员的工作负荷。例如，驾驶舱自动化系统为飞行员执行空管指令提供了方便，一些复杂的离场和进近航路、高度限制等飞行作业可通过飞行管理计算机系统的预编程来完成，一定程度上降低了飞行员的工作负荷。进场和着陆飞行操作是一个高负荷工作场景，临时改变的进场或着陆空管指令会迫使飞行员修改预编的程序，从而将飞行员置于潜在危险的高工作负荷之中，或迫使飞行员放弃有利的全自动化功能优势而回到低水平的自动化飞行操作，而某些规定的离场和进近程序的设计也并非完全与驾驶舱自动化飞行操作程序完全兼容，增加了飞行员的工作负荷。

## 2.4 人因工程的途径

针对自动化驾驶舱的人因问题，本文从以下几个方面来概述人因工程的解决方案。

### 2.4.1 "以人为中心的自动化"设计理念

驾驶舱自动化设计理念决定了飞行员与自动化系统之间的功能分配、系统设计、人机界面设计以及飞行操作程序的制定。针对自动化驾驶舱的人因问题，Billings（1997）提出了"以人为中心的自动化"设计理念，强调系统必须为飞行员提供足够的反馈信息，保持飞行员的在环状态，确保飞行员拥有对飞行的最终决策控制权。Endsley（2017）综合以往的研究提出了一个人-自动化监督（HASO）模型，该模型概括了产生自动化人因问题的主要因素，并且为实现有效的"以人为中心的自动化"理念提供了一个人因工程解决方案。

从飞机制造商角度看，波音和空客公司都强调了"以人为中心"的驾驶舱设计理念，但是具体的人机界面设计则不同。空客的自动化驾驶舱设计向飞行员提供了更多的自动化飞行操作功能和较少的自动化系统反馈信息（比如自动飞行状态下静止的侧操纵杆和油门杆），希望降低飞行员的工作负荷。例如，空客飞机的自动化系统限制了飞行员的可控飞行包络面（横滚小于67度、俯仰角小于30度），这是一种"硬保护"的设计思路。而波音的自动化设计更强调飞行员的主动控制权，采用的"软保护"设计允许飞行员在飞行包络面以外操控飞机，以便在应急状态下允许飞行员主动操控飞机来摆脱困境。另外，波音驾驶舱为飞行员提供了余度式多通道反馈信息（比如失速状态下中央手动操纵杆的抖动，自动飞行状态下动态的操纵杆和油门杆）。围绕这两种自动化设计优劣的争议许多年以来一直没有结论。从飞行安全看，两类飞机事故率都极低，说明了实现"以人为中心的自动化"设计理念的重要性，但是人因工程界目前还缺乏足够的飞行事故数据来对两种设计方案做出一个全面的科学评价（许卫，2004）。

### 2.4.2 人机交互和人机界面优化

从人机界面设计来说，航空界一直在朝着"以人为中心的自动化设计"的方向努力。例如，为降低进场和着陆阶段的可控飞行撞地(CFIT)事故率，增加飞行员的垂直状态意识和低能见度下的可操作性，波音开发了垂直状态显示器(VSD)，其中人因工程研究在显示画面格式设计和实验验证方面发挥了重要作用（Mumaw, Boorman et al., 2000）。新一代的飞机驾驶舱（比如波音787等）也推出易操作、整合化的飞行管理计算机系统的人机显控装置（Neville & Dey, 2012）。从长远看，设计出统一的、跨制造商和机种的标准化自动化驾驶舱是非常有必要的，有利于进一步提高飞行安全，并且降低航空公司的飞行员培训和运营成本（FAA，1996）。

人为差错是导致约70%民机重大事故的主要原因或原因之一，其中由驾驶舱设计引发的飞行员人为差错占有相当大的比例（FAA，2004），航空界和人因工程界正在通过驾驶舱人机界面优化设计和适航测试认证等手段来减少由设计引发的飞行员人为差错（许为，陈勇，2014）。

研究者还从优化人-自动化交互的角度来解决相应的人因问题。例如，针对目前的自动化驾驶舱人机界面没有提供有效信息来支持飞行员应对非正常操作场景的的自适应认知能力和行为，Ackerman等（2017）的研究提出优化自动化情景意识的信息显示来修复人-自动化交互中的人因问题，Calhoun（2021）的近期研究则提出采用人类操作员启动的适应性人-自动化交互机制有助于提升操作员的情境意识和人工干预的有效性，而不是采用传统的系统分配的人-自动化交互机制。

### 2.4.3 适航认证

美国航空总局（FAA）在1999年启动了一项针对自动化驾驶舱人因问题的研究，结果表明FAA适航条款中存在35项与人因有关的问题，这些问题涉及到如何在适航要求中充分考虑飞行员能力、如何有效地支持飞行员作业绩效以及如何有效管理人为差错等方面（FAA，2004）；FAA 25部对人因方面的适航要求是按照"以系统设备和功能为导向"的方式在相关的条款中作出一般性和定性的要求。该

研究认为局部地修改 25 部的部分条款无法系统地解决所存在的人因问题，有必要按照"以飞行员为中心"的方式增补一项新条款（CS25.1302）来系统地解决这些问题（FAA,2013）。25.1302 新条款将整个驾驶舱中与飞行员飞行任务相关的设备和功能视为一个整合的人机交互系统，以飞行员任务为导向，以能否支持飞行员有效和安全地完成规定的飞行任务（作业绩效）为目标，规定了这些设备和功能的设计必须与飞行员的能力相匹配，从而能够有效地支持飞行员作业绩效和人为差错管理，并且最大限度地减少飞行员人为差错等人因问题（罗青，2013）。欧洲航空安全局(EASA)和 FAA 分别于 2007 年和 2013 年正式将该新条款（CS25.1302）纳入适航认证要求（EASE,2007；FAA,2013），中国商飞也已将 25.1302 条款纳入 C919 型号的适航认证中（党亚斌，2012）。

### 2.4.4 人因工程的挑战

Strauch（2017）的研究表明，尽管航空界和人因工程界都在致力于改进自动化驾驶舱的人因问题，但是经典的"自动化的讽刺"现象在30多年后仍然没有得到彻底的解决，波音737 MAX飞机在2018、2019年（狮航和埃塞俄比亚航空）连续发生的与驾驶舱自动化系统有关的两起重大事故说明了这一点（NTSB，2019）。作者多年前的研究表明实现和落实人因工程的研究成果仍然面临着挑战（Xu,2007）。

针对自动化驾驶舱的人因问题，Xu（2007）采用 Rasmussen（1985）的抽象层次结构（AH）认知工程模型，开展了基于大数据的建模分析。这些数据包括：（1）飞行员问卷调查。调查数据代表了在驾驶舱自动化操作方面具备不同航线飞行经验的全球 5000 多名飞行员（BASI，1998）；（2）飞行事故报告。航空安全报告系统（ASRS）数据库中所记录的与驾驶舱自动化飞行直接相关的 105 个报告；(3)飞行模拟器实验。在波音 747-400 飞行模拟器上基于眼动追踪、飞行员自动化监控行为、自动化飞行绩效的实验数据（Sarter, Wickens et al., 2003）；（4）飞行员心理模型测试。飞行员在完成 747-400 飞行模拟器实验以后的结构式访谈数据（Mumaw, Sarter et al., 2000）。该建模分析研究表明，飞行员掌握了驾驶舱自动化操作的基本知识和技能，飞行员失误主要发生在执行与规则式和知识式行为相关的自动化飞行操作时；有些飞行员缺乏一个完整有效的自动化飞行操作的心理模式；自动化人机界面未能有效地呈现与自动化飞行目标相关的信息；飞行员没有足够的能有效应对应急状态的自动化飞行操作程序。该研究建议，改进飞行员培训、飞行手册内容、自动化飞行操作程序，并且进一步优化基于"以人为中心"理念的驾驶舱人机界面。由于驾驶舱人机界面设计改进费时，并且不会影响现有机队，因此研究建议优先考虑飞行员培训、飞行操纵手册和飞行操作程序方面的工作。

针对有效的飞行员自动化培训，采用 Vicente（1999）认知工作分析方法对飞行员自动化培训内容的进一步分析表明，由于缺乏对驾驶舱自动化领域知识的有效和完整的表征，许多飞行员培训内容不能有效地帮助飞行员建立起一个准确和完整的自动化系统的心理模型，也不利于有效地培养飞行员的知识-驱动式监控能力和维持良好的情景意识。这些原因导致个别飞行员遇到复杂操作情景和异常状态时，有时不能有效地在整个领域知识空间内搜索或推理出符合当前飞行场景的自动化飞行操作（Mumaw, Boorman et al., 2000; Xu, 2007）。

导致波音 737 MAX 飞机在 2018、2019 年两起飞机解体事故的发生有多方面的原因，其中包括制造商管理层、自动化系统设计（机动特性增稳系统，MCAS）、人机界面设计（告警信息等）、适航认证、FAA 监督、安全文化、飞行员培训以及地面维修等（Cusumano, 2020; The U.S. House Committee, 2020）。从人因工程角度分析，这两起 737 MAX 飞机事故至少包括以下三个方面的原因："以人为中心自动化"设计理念的实现（如何保证任何时候飞行员拥有对自动化系统的最终决策权，如何保证飞行员在应急状态下能够快速有效地中断自动化系统来人工接管飞行操控），人机界面设计优化（如何在现有驾驶舱告警人机界面中有效地整合新设备的告警信息，从而为飞行员提供足够的情景意识，避免"自动化惊奇"现象等），必要的飞行员培训和飞行手册内容更新（如何保证飞行员拥有一个准确和完整的自动化系统的心理模型）。

由此可见，展望今后自动化驾驶舱的人因工程工作，人因工程界要继续推动以往研究成果在实际应用中的落实；另一方面，人因工程界需要并且也一直在继续开展自动化驾驶舱的研究。例如，在驾驶舱显示界面设计方面，人因工程界研究如何优化信息显示来降低"自动化惊奇"的可能性(Dehais et al.，2015)；研究如何通过以人为中心的自动化显示设计和训练将飞行员的非适度自动化信任和依赖调整到适度状态（王新野等，2017；Manzey et al.，2012）；研究如何通过有效的自动化监控策略、飞行员培训、心理模型构建来帮助飞行员应对非正常飞行场景（Billman et al.，2020）。

最后，智能自主化新技术正在进入人们的日常工作和生活，自动化驾驶舱也不例外。在智能时代，人因工程的挑战与机遇并存：既要在引进智能新技术的同时吸取自动化驾驶舱人因问题的教训，同时要充分发挥智能新技术的优势，找到能够进一步帮助解决自动化驾驶舱人因问题的有效方案。

## 3. 智能化飞机驾驶舱

### 3.1 基本概念

从某种意义上讲，机载自动化技术辅助飞行员的能力达到了瓶颈，在一些特殊操作环境中（比如设计无法预料的非正常操作场景），飞行员需要智能化、更高水平的"自动化"技术的支持。从技术层面上来说，不同于自动化技术，智能技术拥有独特的自主化（autonomy）特征。带有自主化特征的智能系统具有一定程度的认知学习、自适应等能力，在特定的场景下可以自主地完成独立于人工干预、甚至设计事先无法预料的一些特定任务（详见本系列文章之五《五论以用户为中心的设计：从自动化到智能时代的自主化以及自动驾驶车》；许为，2020）。自动化和自主化之间的区别不是在自动化水平上递进的关系，有没有基于智能技术的认知学习、自适应、独立执行等自主化能力是两者之间本质上的区别，智能系统借助于一定的算法、机器学习和大数据训练等手段，可以在一定的场景中自主地完成以往自动化技术所不能完成的任务（Madni & Madni，2018；Kaber，2018）。

如前所述，现有飞机自动化驾驶舱系统是基于自动化技术，其运转依赖于固定的逻辑规则和算法来执行定义好的任务（比如通过飞行管理计算机系统的编程），并产生确定的操作结果，当出现设计无法预料的非正常飞行场景时，需要飞行员通过手动飞行操纵来执行人工干预。从理论上来说，基于智能技术的飞机智能机载系统有可能克服自动化技术的局限性。例如，现有自动化飞行系统不能应对事先设计中无法预测的飞机故障或者非正常飞行场景，通过采用机器学习、大数据（基于大量经验丰富的飞行员操作数据以及相关的飞行参数和操作场景的知识库）等技术，训练并建立能够应对故障和非正常飞行场景模式的智能飞行系统，从而弥补个体飞行员能力和知识的局限，安全有效地摆脱一些飞行操作困境。

针对未来驾驶舱机载智能系统的功能，目前还没有具体完整的解决方案，但是研究者们基于智能技术的优势、自动化技术的局限性、飞行安全和航空公司等需求，展望了未来智能驾驶舱的发展（严林芳等，2017；杨志刚,张炯等，2021；吴文海等，2016）。例如，通过机载智能辅助驾驶技术来实现正常场景和标准飞行流程下"门到门"的全自动驾驶(包括场面滑行、起飞着陆)，并且具备检查单等相关飞行准备工作的全自动执行能力；通过机载智能决策辅助技术为飞行员提供全飞行阶段(正常、非正常)飞行操作的决策支持，以及提供指引以及信息融合(含空管、交通服务等)显示、机组告警系统信息自动关联、系统故障分析与预测等能力；利用智能语音识别技术开发出更加有效的机载人机交互界面；基于知识图谱和结构化数据以及决策树等技术，形成具有飞行驾驶决策能力的数字模型。机载智能系统应该能够感知飞行运营信息和飞机状态（实时动态获取飞机内外部信息）、进行实时获取、分析和预测（天气、地形等）、进行记忆学习（空天一体化网络进行云存储，基于机器学习获取新知识）、自主控制与规划（飞行操作、航路规划、飞行决策等）、装备智能交互技术(触摸屏、组合视景、民用飞机头戴显示、语音识别等)以及进行空地一体化智能维护管理等（严林芳等，2017；吴文海等，2016）。

国际上，欧洲航空安全局(EASA)(2020)发布了《AI 路线图》，波音和空客等企业正积极探索 AI 技术在航空领域的应用。中国商飞在 2020 年发布了"有人监督模式下的大型客机自主飞行技术研究"的技术指南（杨志刚等，2021）。

## 3.2 人因工程初步研究
### 3.2.1 "以人为中心"的智能化驾驶舱

智能自主化技术并非十全十美,它具有"双刃剑"效应(许为,2020)。一方面,智能系统利用大数据、AI 深度学习等技术可以整合大量的专家知识,主动地帮助人类操作员在非正常场景中解决以往单人知识所不能解决的问题,这是传统自动化技术无法实现的;另一方面,如果智能系统设计中不遵循"以人为中心 AI"的理念,不保证人类拥有对系统的最终决策控制权,智能系统自主独立执行和不确定输出等自主化特征有可能带来飞行安全的隐患(许为,葛列众,2020)。

人因工程界针对智能自主化技术已经开展研究。Endsley(2017)认为"自动化的讽刺"效应也会在智能自主化技术应用中出现。在智能自主化系统中,随着单个功能"自动化"水平的提高、整体系统的自主化以及可靠性的增加,操作员对这些功能的关注度和情景意识可能会降低,在应急状态下出现"人在闭环外"的可能性也会增加。近几年发生了许多起基于智能自主化技术的自动驾驶车致命事故,人因分析表明导致这些事故发生的原因包括:人机界面设计问题、情景意识下降、自动化模式混淆、"人在闭环外"、低参与度、过度依赖或信任自主化等,而这些问题正是以往人-自动化交互中的典型人因问题(Navarro,2018;NHTSA,2018;Endsley 2017;Xu,2020)。

另外,具有学习等认知能力的自主化系统随着在不同环境中的使用会不断提升自身的能力,其操作结果具有潜在的不确定性,因此自主化技术有可能比自动化技术给人类操作员带来更强烈的"自动化惊奇"体验(Sarter & Woods,1995)。这些效应可能进一步放大以上这些人因问题的影响程度,这种现象被称为"lumberjack"效应 (Onnasch et al. 2014)。研究还表明,与自动化技术相比,自主化技术还可能会导致操作员高度情绪化的反应,一些社会因素更容易对操作员的心理产生影响 (Clare,et al.,2015; de Visser,et al.,2018)。

由此可见,如同自动化技术在飞机驾驶舱应用的初期,智能化飞机驾驶舱的研发需要人因工程的早期介入和解决方案。正如本系列文章《四论以用户为中心的设计:以人为中心的人工智能》中所提出的"以人为中心的 AI"设计理念(许为,2019),针对民用航空领域的高风险性特征,"以人为中心设计"的理念强调人应该是在任何条件下、任何时候拥有对系统的最终决策控制权。

针对航空智能系统研发,采用"以人为中心"的理念目前已经基本达到了共识(例如,Parnell et al.,2021;EASA 2020;许为,陈勇,董文俊等,2021)。美国国家研究委员会(NRC)在 2014 年发布的《民航自主化研究:迈向飞行新时代》中强调,智能系统的操作需要人的参与,要充分考虑人与智能系统之间的角色、职责和工作量的有效分配,人与智能系统是协作伙伴的关系。EASA 发布的《AI 路线图》的副标题("以人为中心的航空 AI 途径")就明确了智能系统开发的理念(EASA,2020)。中国商飞强调开发智能系统要以人(飞行员)为核心,一切智能化功能均应该围绕飞行员操作和决策需求开发(杨志刚等,2021)。欧洲人因和工效研究院(CIEHF)(2020)在《未来驾驶舱技术的以用户为中心的设计和评估》的白皮书中强调,随着 AI 技术的实施,未来 30 年航空领域将发生许多变化,但是人类在系统控制和决策中仍将发挥关键作用;机载智能系统的设计应该"以飞行员为中心";机载智能系统是提高飞行员情景意识、规划和决策能力的"智能助手",但不是完全取代人类操作员。

### 3.2.2 基于智能技术的初步探索

针对智能驾驶舱的探索研究多年前就已经展开,尽管当时的智能技术还不成熟,但是人因工程专业人士与其他学科合作提出了一些概念,并且取得了一些初步结果。例如,飞行员助理(Smith & Broadwell,1986)、旋翼机飞行员助理(Miller & Hannen,1999)和 CAMA(Onken,1999)。其中,Gerlach & Onken(1995)基于专家知识开发了民航驾驶舱智能决策辅系统(CASSY),该系统包括飞行员模型、目标冲突评估、飞行员意图和差错识别等功能,在各种飞行场景中通过对飞行状等评估和规划来协助飞行员,初步测试证明该系统是有效的。近几年 AI 技术的快速发展也进一步推动了这方面的研究。

例如，Vemuru 等（2019）的研究开发了一个基于强化学习代理技术的智能系统，该系统通过在模拟器中感应飞行员的心理反应和飞行路径来学习飞行技能，并且与飞行员形成一种协同合作关系，经过训练的智能代理可以在飞行的各个阶段协助飞行员作业。

在国内，中国商飞开展了针对智能化驾驶舱的初步研究（大漠，2016）。该研究对未来智能化驾驶舱提出了简洁、智能、互联、安全四大理念，将智能决策、主动优化、虚拟现实、触摸控制、机载互联、语音控制等技术引入民机驾驶舱中，提出了全玻璃化驾驶舱概念,并且设计了拥有全新人机显控界面布局的展示模拟驾驶舱。国外的一些公司和研究机构也正在开发机载智能系统等（郭建奇，2020），例如，泰雷兹（Thales）（2019）公布了正在研发的一些技术，其中包括为下一代 FlytX 航空电子设备套件之一的一个虚拟助手，该助手可为飞行员提供语音人机交互和对飞行员实现一定程度上飞行员意图识别；Daedalean 在开发首个航空自动驾驶仪系统中采用了深卷积前馈神经网络的AI技术。

### 3.2.3 智能自主化水平

如何根据智能技术的自主化特征来有效定义智能系统的自主化水平（autonomous level）是一个重要的人因工程研究问题。它不仅涉及到如何区分自动化与自主化之间的特征，并且也涉及到如何有效地划分智能自主化的等级水平，从而在系统设计中达到人与智能系统的最佳匹配和协同合作。在自动化领域，自动化水平的定义直接影响到系统设计中系统功能和控制权在人与自动化系统之间的有效分配。Parasuraman 等人（2000）从人类信息加工角度提出了自动化的"4 阶段理论"（信息获取、选择和过滤,信息整合，行动选择，控制和行动执行），Sheridan & Verplank（1978）将自动化划分为 10 个水平。已有研究者认为传统的自动化水平划分并没有充分考虑到智能自主化技术的特征，因为这些自动化水平的划分主要是从人机功能分配的角度出发，并没有考虑到智能技术的自主化特征，例如在特定场景下所具有的一定程度的认知学习、自适应、自治（self-governance）、自主执行等自主化能力（Kaber, 2018；Bradshaw et al., 2013；Kistan et al., 2018；许为，2020）。因此这些传统的自动化水平定义不适合于智能自主化技术。

航空领域目前还没有被广泛接受的智能自主化等级划分方法。有研究者强调人的参与程度和人类与智能自主系统之间的合作伙伴关系，将自主化分为人操作、人授权、人监督和完全自主 4 个等级(卢新来等，2021)。经过二十多年的研究，虽然目前依然没有建立起具有实际操作意义的自主等级划分方法，但是人们对智能自主化以及自主化等级等概念的认知正在不断深化。例如，EASE（2020）在 AI 路线图中提出了对应于智能自主化层次上的三种主要场景：人在回路中（HITL）、人在回路上（HOTL）和人在控制（HIC）。在 1 级（人在回路）中，驾驶舱机载智能系统扮演"助理"角色，协助机组人员操作，人类飞行员起着主控作用；在 2 级（人在回路上），机载智能系统开展与飞行员之间的协作合作，机组人员仍然全权负责飞行和监控；在 3 级（人在控制）中，机载智能系统拥有更高的自主化层次，在正常操作运行中，智能自主系统将不要求人的操作，但人负责监督并且拥有最终的主控权。由此可见，对智能自主化程度的划分正越来越多地考虑"以人为中心"的设计理念。

### 3.2.4 人-智能系统的协同合作

人与智能系统的协同合作是另一个重要的人因工程研究课题。不同于作为"辅助工具"的自动化系统，智能自主系统可以成为与人类合作的"队友"，分享任务和操控权，形成"人机组队"（human-machine teaming）式合作的新型人机关系，或者称为"人-自主化组队"（human-autonomy teaming）式合作（Kaber, 2018；许为，葛列众，高在峰，2021）。在民航领域，研究者已经开展了针对"人-自主化组队"的一系列人因工程。初步研究表明，通过培养适当人-自主化组队合作，人机之间会增加互信，提高人机系统的绩效（Ho et al., 2017; Tokadli et al., 2021）。

在针对大型商用飞机单人飞行操作（single pilot operations，SPO）的研究中，研究者采用"人-自主化组队"式合作的概念希望在飞机驾驶舱引进一个"智能副驾驶"来承担起与飞行员合作的队友角

色，形成类似于双乘员驾驶舱的机组合作关系。例如，Tokadli 等人(2021)采用一个"剧本委托界面"（PDI）来评估 SPO 驾驶舱中的"人-自主化组队"式合作。该系统是一个基于领域知识库和决策-行为架构的智能自主系统，在一些设计无法预料的操作场景中可以辅助操作员。初步实验表明 PDI 有助于飞行员与该自主系统的合作。Lim 等人(2017) 提出的 SPO "虚拟飞行员助理"（VPA）系统架构包括推理模型、不确定性分析模型、认知知识模型等，目的是通过 SPO 飞行员与智能系统之间的协作来降低工作负荷。Brandt et al.(2017) 采用"人-自主化组队"式合作理念开展了针对地面站操作员执行 SPO 飞行跟踪任务的评估，该研究采用了基于自动推荐系统的自主约束飞行计划器（ACFP）系统，目的是通过该系统与地面站操作员的协作关系来支持地面站操作员在非正常场景中的快速决策。初步模拟实验结果表明，与没有 ACFP 的地面站相比，参与者认为 ACFP 提供了足够的情景意识，降低了工作负荷。

### 3.3 人因工程研究展望

展望航空智能技术应用的前景，EASA(2020)的 AI 路线图基于"以人为中心"理念提出了三阶段目标：第一阶段（2019-2024）利用智能技术协助人类操作和增强人类能力；第二阶段（2024-2028)开发基于人机协同合作的智能系统；第三阶段(2029-2035)开发高水平自主化的智能系统，同时确保人是最终决控者。CIHFE(2020)从人因工程的角度提出了一份航空领域 AI 发展路线图，中期目标（2025－2035)强调开发智能数字助理和可解释 AI 技术，保证智能系统能够为机组人员提供有效的决策建议；长远目标(2035－2050)是建立智能化空中交通系统，实现智能化空中交通。NASA（2019）《航空战略实施规划》中定义了近远期目标。近期目标（2015-2025 年）是引入具有有限的自主化、执行功能级目标的航空系统；中期目标（2025-2035 年）是引入具有灵活的自主化、可信任、可实现任务级目标的航空系统，并且人和系统合作完成任务目标；远期目标（2035 年后）是引入具有自主性、可实现政策层面目标的分布式协作航空系统。NASA 规划还强调开展针对人与智能系统合作、人-自主化组队合作的研究。可见，以上这些发展路线图都强调了人在系统中的作用以及人与智能系统的协同合作。

针对智能驾驶舱人因工程的研究目前处于起步阶段，本文针对今后人因工程研究和设计总体工作思路提出以下一些初步建议。首先，采纳"以人为中心"的理念来研发智能系统，将人放在系统研发的中心位置考虑，发挥人类与机器智能间的优势互补，实现"人在环"的系统设计，保证人拥有对飞机的最终操控权。人因工程要保证智能系统中有效的人机功能分配，即人类从事战略、规划和决策性任务，而系统则负责操作性任务，保证人与智能系统之间的最佳人机匹配和协同合作。

其次，开展机载人-智能系统交互、协同合作的基础研究。针对人机协同合作计算模型和定量评估，构建和实验验证基于"人-自主化组队"合作的人机交互和决策模式及设计概念。采用多学科方法开展"人-自主化组队"研究，例如，采用协同认知系统理论（Hollnagel & Woods，2005），探索如何利用人类生物智能和机器智能在不同程度上的深度组合以及互补来支持基于人机组队合作的自主化创新设计；开展人与智能系统之间人机互信、情景意识分享、决策分享、控制分享的认知和计算建模研究，为系统设计提供人因工程解决方案。

第三，从航空安全和智能技术成熟度看，目前大型商用飞机驾驶舱机载系统设备升级不可能全部采用智能自主化系统，应该是一种智能化与自动化的组合系统。人因工程应该根据人机功能分配、技术的各自特点，发挥自动化和自主化技术的优势互补，在人、自动化以及自主化三者之间找到最佳的"以人为中心"的人因工程解决方案。例如，飞行员的尖峰工作负荷阶段对应的飞行任务可以是一个智能飞行技术融合的合适切入点，对当前航空技术的发展具有实际意义，也不会对目前的航空技术体系和驾驶模式产生大影响，有利于通过技术迭代来积累智能化技术融合的经验（杨志刚等，2021）。

最后，在人与智能系统交互和人机界面设计方面，要吸取自动化驾驶舱人因问题的教训，优化人机界面设计，考虑采用创新方法设计人-智能自主化交互的人机界面，比如基于"人-自主化组队"式合作的人机交互设计，开发驾驶舱智能化人机交互界面。另外，智能飞行技术的引进将飞行员的角色从直接

飞行操纵员转向飞机资源管理者，人机界面设计必须保证飞行员有足够的情景意识，在智能技术失效时驾驶员具备时刻接管飞机的能力。

## 4. 个案分析：大型商用飞机单人飞行操作

### 4.1 基本概念及"以人为中心"的设计理念

过去 50 多年中，技术发展推动了大型商用飞机驾驶舱机组人员逐步递减(de-crewing)的趋势，从最初 5 名机组人员到目前机长和副驾驶 2 人配置，这种递减情况在不久的将来还会持续下去。作为新一代商用飞机发展核心技术之一，目前国内外民航界正在积极探索和研发大型商用飞机"单一飞行员驾驶"（SPO）模式。SPO 指在大型民机驾驶舱中仅配置一名飞行员，借助提升的机载设备或者远程地面站操作员的支持（或者两者组合），能够在各种飞行场景中安全有效地完成航线飞行任务，并且达到不低于目前双乘员驾驶模式的飞行安全水平(Comerford et al., 2013)。SPO 会导致一场航空运输革命，在满足安全性条件下，SPO 可以带来减少飞行员数量、提升经济性、减少驾驶舱资源配置、缩小驾驶舱空间和减轻飞机重量等方面的好处。

美国、欧盟、Boeing、Airbus 等飞机制造商正在开展 SPO 研发(Comerford et al., 2013)，国内针对 SPO 技术方案和系统架构的一些研究工作也开始起步(王淼等，2020)。初期的 SPO 研究主要涉及两种 SPO 总体方案（Matessa et al., 2017）。一种是驾驶舱机载设备更新方案（简称驾驶舱方案），主要通过提升现有驾驶舱机载自动化系统或引进机载智能自主化系统来替代现有人类副驾驶的部份职责，SPO 飞行操作基本上不依赖于地面支持；另一种是远程地面站飞行支持方案（简称地面站方案），该方案具有分布式机组的设计概念，即将现有副驾驶的部份职责从空中移到了地面。随着研究的展开，研究者开始考虑第三种方案："SPO 驾驶舱飞行员 + 驾驶舱机载设备提升 + 地面站飞行支持"协同实现的 SPO 组合式模式（见图 3）。

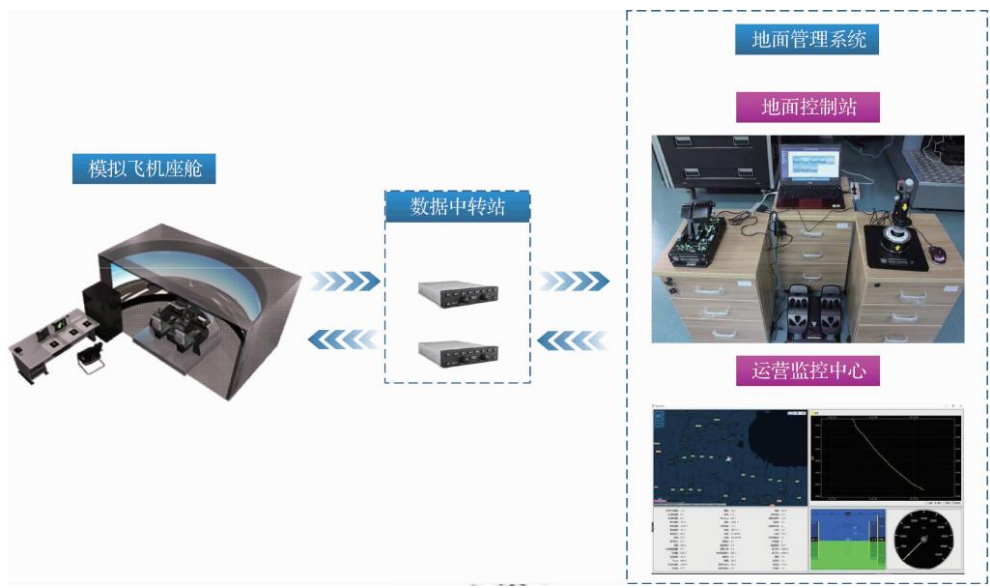

图 3 "驾驶舱 + 地面站"SPO 组合方案演示和验证系统架构（王淼等，2020)

人因工程预备研究在飞机型号研发中起着无可替代的重要作用(许为，陈勇，2012)，SPO 研发也不例外。在美国以 NASA 为主开展了一系列 SPO 人因工程研究，欧洲和澳大利亚等地的科研院校也开展了 SPO 人因工程研究，国内这方面的工作尚未正式启动。人因工程界强调实现 SPO 的最大障碍不是技术本身，

而是如何遵循"以人为中心设计"的理念，合理利用技术，研发出一个有效支持SPO飞行安全的人因工程解决方案(Harris,2007；许为，陈勇，董文俊等，2021)。

### 4.2 人因工程初步研究

2012 年 的NASA SPO 技术交流会确定了人因工程研究的5个重点领域：SPO设计方案、设备系统更新、人员交流与沟通、飞行员失能、适航认证。目前的人因工程研究主要集中在前三个方面，并取得了一些阶段性成果(详见许为，陈勇，董文俊等，2021)。实现SPO的必要条件之一是更新现有驾驶舱机载设备系统，争议的焦点之一是提升现有驾驶舱机载自动化系统还是引进机载智能自主化系统。

许多研究者出于对SPO整体设计复杂性的担忧，只是在一个"舒适区"内提出一些针对现有系统局部改进的方案，虽然这种"演变"式路径对系统开发和适航认证成本很低，但是很难从根本上解决人-自动化交互中的问题。从长远来看，这种方法可能带来飞行安全、人因工程、系统可扩展性等方面的问题。Sprengart 等人(2018)认为，离开"舒适区"是寻找SPO系统方案的必要条件，只有这样才能将SPO飞行员与机载系统（自动化或智能自主化）在人-机、空-地的飞行协同操控和决策等方面达到最佳的匹配。Neis 等人(2018)建议，SPO系统方案的设计应该回到原点，将人放回中心，让技术适应人，最终达到人机关系的最佳匹配。因此，SPO应该是优化驾驶舱人机交互设计的一个新机遇，有利于解决"历史遗留"的人-自动化交互的人因问题。

研究者希望在SPO驾驶舱引进一个"智能副驾驶"来承担起与SPO飞行员合作的队友角色，形成类似于双乘员驾驶舱的机组合作关系来解决SPO的一些挑战。例如，Lim 等人(2017)提出的SPO"虚拟飞行员助理"等概念。在NASA、FAA和Rockwell 早期合作的一项SPO模拟舱实验研究报告中，该报告建议SPO的技术干预方案不应该仅仅是提升现有驾驶舱自动化系统，还要考虑引进新的智能自主化系统(Bailey et al.，2017)。尽管这方面的整体技术和人因工程研究尚不成熟，但是在民机机载设备领域已有一些正在研发的智能子系统(详见许为，陈勇，董文俊等，2021)。例如，智能化推荐检查表及状态传感系统，机载人机语音交互，智能化空中交通防撞系统，应对故障模式的智能飞行系统，可穿戴智能设备。这些研发有助于为SPO机载智能自主系统和SPO地面支持站的研发提供支持。

根据人因工程研究、机载自动化技术和智能技术可行性，许为，陈勇，董文俊等人（2021）建议SPO驾驶舱机载系统的提升需要采用"自动化 + 自主化"的组合式方案，即根据场景复杂性选用技术，利用两种技术的优势互补来获取最大的安全保证。SPO研发需要采纳"以人为中心设计"的理念，将人放在系统研发的中心位置考虑，发挥人类与机器智能间的优势互补，实现"人在环"的系统设计，保证人拥有对SPO飞机的最终操控权。针对驾驶舱设备升级，需要开展人机交互、协作和决策的人因工程研究，为SPO空地人-人与人-机交互、协同和决策的系统整体设计提供方案。例如，基于智能技术，构建和实验验证基于"人-自主化组队"合作的人-自主化之间的人机交互和决策模式及设计概念；在以往研究基础上，构建和实验验证SPO飞行员与自动化之间的人机交互决策模式和设计方案。

### 4.3 智能SPO系统的人因工程分析

本文采用上一篇系列文章《六论以用户为中心的设计：智能人机交互的人因工程途径》中作者所提出的智能人机交互（iHCI）人因工程模型来分析SPO模式（许为，2022）。iHCI 人因工程模型将人类操作员和智能系统（智能代理，intelligent agent）均视作为能够完成一定认知信息加工任务的认知体，从而一个iHCI系统可以被视为一个协同认知系统（Hollnagel & Woods，2005）。iHCI 人因工程框架强调人机协同合作、人机双向主动式状态识别、人类智能与机器智能的互补性以及人机合作式认知界面等特征。基于iHCI 人因工程模型，SPO驾驶舱飞行员与机载智能系统之间的交互就是一种智能人机交互，SPO驾驶舱机载人机智能系统就是两个认知体（SPO飞行员，机载智能系统）协同合作的一个认知协同系统。

进一步地，"SPO驾驶舱飞行员 + 驾驶舱机载智能系统 + 地面站飞行支持"组合SPO方案强调人-机与空-地的飞行协同合作，该SPO模式整体系统的安全运营不仅仅取决于单机中的"SPO驾驶舱飞行员

+ 机载智能系统"认知协同系统，还取决于来自"地面站飞行支持"、"智能空中交通系统架构"、"智能化社会技术系统"认知协同系统的支持。所有这些认知协同系统形成了一个大型商用飞机智能 SPO 协同认知生态系统(见图 4)。我们采用协同认知生态系统的框架来进一步分析 SPO 的人因工程解决方案。

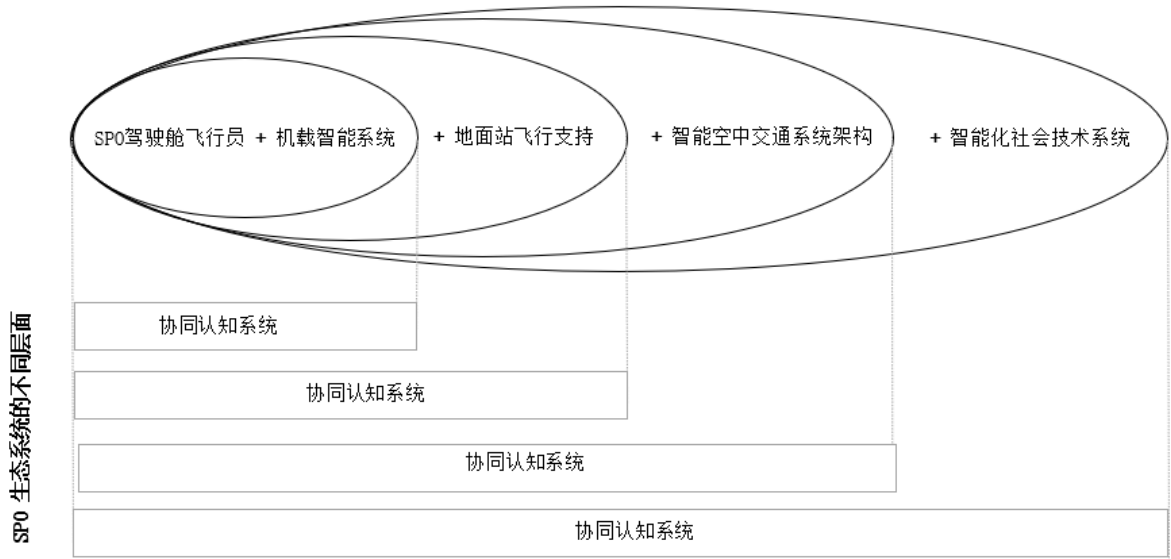

**图 4 基于"驾驶舱飞行员 + 驾驶舱机载智能系统 + 地面站飞行支持"的智能 SPO 协同认知生态系统**

协同认知生态系统是一个多层次的系统架构，一个子系统的协同认知系统可以是整体协同认知生态系统架构中的某一层次。图 4 示意了智能化 SPO 协同认知生态系统的多层次架构，表 1 概括了该协同认知生态系统各层次的系统组成部分。其中，认知协同系统的范围和边界条件是相对的，取决于分析的目的。

**表 1 智能 SPO 协同认知生态系统中各层次的主要组成部分**

| 系统组成<br>SPO 协同认知生态<br>系统的架构层次 | SPO 驾驶舱单人飞行员，机载智能系统（环境和驾驶员状态感应系统、机载智能体等） | 地面站智能系统，地面站支持人员，云技术，5G 等 | 智能空中交通信号系统，空中交通规则和法律，智能空中交通指挥系统（ATM）统等 | 公众，适航论证当局，航空公司，飞机制造公司等 |
|---|---|---|---|---|
| SPO 驾驶舱飞行员 + 机载智能系统 | √ | | | |
| SPO 驾驶舱飞行员 + 机载智能系统 + **地面站飞行支持** | √ | √ | | |
| SPO 驾驶舱飞行员 + 机载智能系统 + 地面站飞行支持 + **智能空中交通系统架构** | √ | √ | √ | |
| SPO 驾驶舱飞行员 + 机载智能系统 + 地面站飞行支持 + 智能空中交通系统架构 + **智能化社会技术系统** | √ | √ | √ | √ |

构建一个"SPO 驾驶舱飞行员 + 机载智能系统 + 地面站飞行支持"SPO 模式涉及到人-机（机载系统）之间以及人-人（地面站）空地之间的协同操控和决策、人-自动化/自主化交互等一系列问题，对安全飞行操作提出了许多新的要求，需要人因工程的支持。该 SPO 模式的实现需要优化人-机、空-地人员之间的功能分配，严格评估在正常以及应急 SPO 飞行场景中各类人员的工作负荷、人机交互决策模式、空-地协同的飞行操控和决策等因素。例如，一方面，SPO 改变了驾驶舱飞行员的认知决策模式，SPO 避免了现有双乘员驾驶舱中双人飞行员之间潜在的认知决策冲突，有助于提升决策效率；另一方面，SPO 驾驶舱飞行员将更多地依赖于个人知识、人-机之间以及人-人空地之间的沟通和协同操控，因此 SPO 的飞行操控和决策模式可能变得更为复杂。另外，SPO 飞行操控权的授权管理和权限分配可发生在人-机（机载、地面站系统）或者人-人（驾驶舱与地面站）之间，这个过程可能会出现飞行操控决策和权限分配方面的冲突。

相对传统的人因工程研究思路，协同认知生态系统框架表现出它的优势。传统的人因工程研究思路一般注重于单机人机系统（"SPO 驾驶舱飞行员 + 机载智能系统"）的问题，但这只是整个智能 SPO 协同认知生态系统中的一个协同认知子系统，影响该子系统绩效和安全的因素不仅仅是 SPO 驾驶舱中两个认知体之间的协同合作，还包括整个生态系统中其他层次上协同认知系统中各种认知体之间的协同合作。如图 4 和表 1 所示，如果将 SPO 解决方案的分析边界逐步扩展到地面站飞行支持、智能空中交通系统架构以及智能化社会技术系统，实现 SPO 整体系统的优化设计则取决于各种认知体之间的协同合作，其中包括地面站支持人员的协同合作、智能空中交通指挥系统操作员的协同合作、公众对 SPO 的认知和接受度、SPO 适航认证规范、航空公司运营等。因此，只有从协同认知生态系统的系统化角度出发，综合考虑整个 SPO 协同认知生态系统内各种认知体之间的协同合作，才能为 SPO 系统的整体优化设计提供一个完整的人因工程解决方案。

### 4.4 智能 SPO 系统的人因工程初步解决方案

基于以上分析和讨论，本文对今后"SPO 驾驶舱飞行员 + 驾驶舱机载智能系统 + 地面站飞行支持"SPO 模式的研究提出以下初步的建议。

#### 4.4.1 "以人为中心 AI"的设计理念

强调"以人为中心 AI"的设计理念。SPO 系统设计要充分考虑人的需要、能力、潜力和极限，利用人已有的经验、知识和技能。例如，机载智能系统不应该完全取代飞行员，应该有针对性地发挥和增强飞行员的潜能（处理异常情况等）以及技能（基本飞行技能等）。现有双乘员驾驶舱的自动化人机交互设计主要是基于"以技术为中心"理念，这种设计将一些飞行操作任务整体地分配给了自动化系统，没有充分考虑飞行员潜能和技能，不能有效地利用飞行员的潜能（处理异常状况等），也浪费了飞行员的基本飞行技能(Billings, 1997)。SPO 设计必须采用"以人为中心"的理念来指导 SPO 人机功能和任务分配，要制定出完整的 SPO 人机功能和任务分配方案。SPO 驾驶舱自动化的提升或者智能化技术的引进会进一步减少手动操控，要考虑如何既能够发挥飞行员潜能和技能，又能保证飞行员"人在环内"的系统设计。

同时，通过有效的余度化设计和操控权权限管理，保证人拥有对 SPO 飞机的最终操控权。例如，除非发生意外状况（飞行员失能等），SPO 驾驶舱飞行员拥有对飞机的最终操控权；在应急状态下，如果飞行员失能发生，机载（自动化研究智能自主化）系统及时找到合适机场，启动应急着陆系统，控操飞机自动着陆；在这过程中，地面站操作员全程实时监控；若综合相关信息和充分证据表明，地面站操作员的介入是必要的，地面站操作员拥有飞机的最终操控权，可随时接管 SPO 飞机安全着陆。

#### 4.4.2 人机协同合作的设计新范式

SPO 智能驾驶舱是由两个认知体（SPO 驾驶舱飞行员，机器认知体/"机器副驾驶"）组成的一个协同认知系统，机载智能系统不仅是支持飞行员的一个辅助工具，而且也是一名与飞行员合作的团队成员

（Hollnagel & Woods, 2005；许为, 葛列众, 2020）。例如，采用"智能副驾驶"系统来承担人类副驾驶的一些职责，人机之间可以分享情景意识、任务、目标和飞行操控权等。

人机协同合作的设计新范式要求重新考虑人和机器在人机系统中各自的最佳角色，要严格定义飞行员与智能自主化系统在飞行操作中应该分别扮演什么角色。例如，未来SPO驾驶舱飞行员的角色是否会从具体航线飞行操作过渡到承担航线规划等"任务管理"的角色；如何根据智能化等级和飞行操作场景来确定人机之间的协同分工合作；研究如何保证在应急状态下飞机操控权在人机之间的快速有效切换，确保人拥有最终控制权。今后的研究要制定飞机操控决策权管理和权限分配，例如，基于定义的分配原则，智能自主技术可实现对SPO飞行操控权限的动态分配，当智能化空中交通防撞系统(ITCAS)在检测到即将发生碰撞时，同时系统检测到飞行员失能或无法及时做出反应时，系统可以自适应调整拥有的权限级别来接管飞行操控（ATI, 2019）。

### 4.4.3 协同认知生态系统的设计视野

SPO系统设计需要从宏观的协同认知生态系统角度出发，任何局限于"SPO驾驶舱飞行员 + 智能机载系统"单机层面的设计方案都无法保证整个协同认知生态系统的优化设计和安全运行。因此，必须开展对整个SPO协同认知生态系统优化设计的研究，其中包括驾驶舱飞行员、机载智能和自动化系统、地面飞行支持站、智能空中交通系统架构以及智能化社会技术系统。

从地面站飞行支持和智能空中交通系统架构看，需要研究不同空间上人-人、设备（地面站与驾驶舱）、人机之间的沟通以及在操控权分享和转移过程中潜在的冲突，要考虑地面站操作员、空中交通管制员和驾驶舱的一体化协同模式（王淼等, 2020；Comerford et al., 2013）。

从智能化社会技术系统角度看，研究要考虑如何获取公众对SPO的认知和信任、飞机制造商与适航当局在SPO飞机适航论证方面的合作、航空公司对SPO研发的参与和支持（地面战飞行支持设备，运营、人力资源等）（许为, 陈勇, 2013）。

### 4.4.4 基于智能自主化特征的设计思路

根据人因工程研究、机载自动化和智能自主化技术的特点，建议SPO驾驶舱设备系统升级考虑采用"自动化 + 自主化" 的组合式方案，并且根据场景复杂性选用技术，利用两种技术的优势互补来获取最大的安全保证。例如，提升现有机载自动化系统来优化面向一般飞行场景的自动飞行模式，引进智能系统来支持面向复杂飞行场景的自主飞行模式（即可独立执行一些设计无法预期的非正常飞行场景）。人因工程要从人机功能分配、工作负荷、人机交互和协同合作等方面出发，优化人-自动化-智能自主化系统三者之间的整合设计，通过实验验证最终的技术方案

针对驾驶舱机载自动化升级设计，系统设计需要避免"自动化讽刺"现象，解决目前双乘员驾驶舱中的人因问题（简化自动驾驶模式，避免"人在环外"等），在人与自动化（自动化水平、人机功能分配、工作负荷等）之间找到一个最佳设计平衡点，保证只有通过严格人因工程实验验证的自动化升级方案才能在SPO方案中被考虑。

### 4.4.5 飞行员失能监控和人机协同控制

针对SPO驾驶舱飞行员失能对飞行安全的影响，作为协同合作的智能机器认知体，SPO驾驶舱必须装备机载感知智能系统来准确监控飞行员状态，一旦发现飞行员进入失能状态，机载系统或者地面站操作员必须快速接管SPO飞机安全着陆。人因工程架构强调人机两个认知体之间双向主动的状态识别，一方面要加强机器认知体对飞行员生理、认知、情感、行为和意图的监测识别和理解；另一方面，加强飞行员对智能系统和环境的情景意识，从而保证人机之间有效的情景意识分享、人机互信、任务和目标分享、决策和控制分享等。

开展针对SPO飞行员失能监测指标的人因工程研究，利用人因工程实验研究来筛选最佳监测指标以及触发告警的最佳阀限值。另外，开展针对SPO飞行员失能的机载监测手段的人因工程研究。例如，监

测系统的人机交互，系统舒服性等（非侵入式测量，远程监控等），监测准确性和预测性（脸部识别，脑电测量等）。

针对从飞行员失能事件发生至 SPO 飞机安全着落期间的安全问题（类似一大型无人飞机），人因工程要提供解决方案，其中包括飞行员失能监控、机载系统（自动化、智能化系统）的自动接管、地面站紧急飞行支持、地面站操作员情景意识和角色转换等。

### 4.4.6  协同合作式人机界面设计

研究协同合作式人机界面设计新范式来有效支持人机协同合作、人机互信、情景意识共享、人机状态识别、控制共享等需求。现有双乘员驾驶舱自动化人机交互设计造成了将飞行员置于"人在环外"的状况，容易降低情景意识，无法迅速有效地处理复杂的意外情况。SPO 为新型驾驶舱人机交互界面的优化设计提供了一个新的机遇，SPO 驾驶舱人机交互设计要基于"以飞行员为中心"的理念，解决现有双乘员驾驶舱人机交互存在的问题。

例如，基于飞行员任务和工作负荷的人机界面动态优化显示方式，简化驾驶舱自动化控制和显示方式；采用创新设计方法（比如基于"人-自主化组队"式合作的交互设计）；采用基于自适应机制的智能人机交互，根据飞行员状态以及场景动态调整人机功能分配。在低负荷操作中鼓励手动操控，保持"人在环"的状态；在高负荷操作中，系统控制飞机，人机界面要突出当前飞行目标参数的显示，使飞行员能够有效执行航线规划或者应急任务。

## 5.  总结

（1）大型商用飞机驾驶舱自动化系统的最初设计基于"以技术为中心"的理念，导致一些人因问题的产生，也给后期改进带来了困难。航空界一直在努力采用"以人为中心自动化"的设计理念不断改进机载人-自动化交互设计，人因工程还面临挑战和还有许多工作要进一步开展。

（2）智能技术为进一步提升大型商用飞机驾驶舱人机交互设计和飞行安全提供了新的机遇。基于"以人为中心 AI"理念的人因工程研究已经初步展开。为进一步提升机载人机交互设计和飞行安全，航空界对驾驶舱机载智能系统研究和应用寄予希望。

（3）根据我们提出的针对智能人机交互的人因工程模型以及协同认知生态系统框架，本文分析了大型商用飞机单人飞行操作模式的人因问题和研究，并且提出了人因工程的初步解决方案。

## 参考文献